\documentstyle[aps]{revtex}

\newcommand{\be}{\begin{equation}}
\newcommand{\ee}{\end{equation}}
\newcommand{\lb}{\left[}
\newcommand{\rb}{\right]}

\begin{document}

\title{Limits on Black Hole Formation from Cosmic String Loops}

\author{Jane H. MacGibbon$^1$\footnote{email: 
macgibbon@snmail.jsc.nasa.gov}, Robert H. Brandenberger$^2$
\footnote[2]{email: rhb@het.brown.edu} and U. F. Wichoski$^2$
\footnote[3]{email: wichoski@het.brown.edu}} 

\smallskip
\address{~\\$^1$Code SN3, NASA Johnson Space 
Center, Houston, TX 77058, USA}

\address{~\\$^2$Physics Department, Brown 
University, Providence, RI 02912, USA.}
 
\maketitle 

\vskip 1.5cm
\begin{abstract}

\noindent 
In theories with cosmic strings, a small fraction of string loops may 
collapse to form black holes. In this Letter, various constraints on 
such models involving black holes are considered. Hawking radiation 
from black holes, gamma and cosmic ray flux limits and constraints 
from the possible formation of stable black hole remnants are 
reanalyzed. The constraints which emerge from these considerations are 
remarkably close to those derived from the normalization of the 
cosmic string model to the cosmic microwave background anisotropies.  
\end{abstract}
\vfill

\setcounter{page}{0}
\thispagestyle{empty}

\vfill 

\noindent BROWN-HET-1084\hfill      July   1997.

\hfill Typeset in REV\TeX

\vfill\eject

\baselineskip 24pt plus 2pt minus 2pt


\section{Introduction} 

Cosmic strings are linear topological defects which are predicted to 
form during phase transitions in the very early Universe in many 
particle physics models of matter (for recent reviews see e.g. 
Refs.\cite{VSrev,HKrev,RBrev}). 
If the scale of symmetry breaking $\eta$ in the theory is of the order 
$10^{16} GeV$ and therefore the mass per unit length $\mu$ of the string 
satisfies $G\mu / c^2 \sim 10^{-6}$, then the strings provide a possible 
mechanism of producing the structure observed today on cosmological 
scales.

The cosmic string theory of structure formation leads to many specific 
predictions for observations. However, many of these predictions 
are almost identical to those of inflation-based models. This is 
one reason for studying other astrophysical constraints on particle 
physics models with strings, in addition to the constraints derived from 
structure formation arguments (we include constraints from cosmic 
microwave background (CMB) anisotropy measurements in the latter 
category). 
The second and maybe even more important reason for studying other 
astrophysical constraints is that strings created in phase 
transitions at symmetry breaking scales much lower than 
$10^{16}GeV$ will contribute negligibly to structure 
formation and the CMB anisotropies and therefore other methods are 
needed to constrain such models. When studying astrophysical constraints, 
it is thus important to keep the scale of symmetry breaking as a free  
parameter.

In this Letter, we shall reconsider the astrophysical constraints which 
arise from the probability that a small subset of string 
loops collapses to form black holes. We will work within the 
``standard" cosmic string model$^{\cite{Zel,V81}}$, according to which 
the network of linear defects formed during a symmetry breaking phase 
transition in the very early Universe quickly reaches a ``scaling" 
solution characterized by having the statistical properties of 
the string distribution independent of time if all lengths are 
scaled to the Hubble radius. 

Strings are either infinite or closed loops. For practical purposes one 
counts as ``long" strings all infinite strings or string loops with 
curvature radius larger than the Hubble radius, and as ``small" loops 
all remaining 
loops. Small string loops are continuously produced 
during the expansion of the Universe by the interaction of long strings. 
Recent numerical simulations$^{\cite{BB,AS,AT}}$ show that there is a 
substantial amount of small-scale structure on the long strings. This 
small-scale structure leads to the production of loops with formation 
radii substantially smaller than the Hubble radius $t$. (In contrast, 
most early work on cosmic strings and structure formation assumed 
that loops were created with radii comparable to the Hubble radius). 
We take the length of a string loop at the time of formation $t'$ to 
be $l = \alpha c t'$ and hence the string loop mass to be 
\be \label{inmass}
m(t^\prime) \, = \alpha c \mu t^\prime \, ,
\ee
where $\alpha$ is a constant which from recent numerical 
work$^{\cite{BB,AS,AT}}$ 
is expected to be much smaller than $1$. Once formed, string loops 
slowly decay by emitting gravitational radiation$^{\cite{VaVil}}$. 
Hence, the string scenario predicts a large stochastic background of 
gravity waves. 

As was shown in the initial work on the string scenario of structure 
formation (see e.g. Refs. \cite{TB,Steb,Sato}), a value of 
$G\mu / c^2 \sim 10^{-6}$ leads to a reasonable amplitude for 
gravitational clustering (as expressed, for example, by the galaxy 
correlation function). The exact amplitude depends 
on $\alpha$, on the average number $\nu$ of long strings crossing 
each Hubble volume, and on the amount of small-scale structure on 
the strings 
(see e.g. \cite{LP95} and references quoted therein). Since cosmic 
strings also lead to CMB anisotropies, the string model can be 
normalized by the recent COBE observations$^{\cite{COBE}}$. This 
normalization is more certain because the calculations involve 
only linear gravitational perturbations. Initial computations of 
CMB anisotropies gave the constraints$^{\cite{BBS}}$ 
$G\mu / c^2 \leq 1.5(\pm0.5) \times 10^{-6}$ and$^{\cite{LP93}}$ 
$G\mu / c^2 \leq 1.7(\pm0.7) \times 10^{-6}$. A more detailed recent 
calculation$^{\cite{ACDKSS}}$ which includes not only the scalar but 
also the vector and tensor modes leads to a 
very similar result $G\mu / c^2 \leq 1.7 \times 10^{-6}$.
\footnote{No error bars are given in \cite{ACDKSS}. However, the 
errors are presumably 
comparable to those of a calculation$^{\cite{ACSSV}}$ using the same 
cosmic string simulation but only scalar modes, namely $\pm 0.35$.} 
Note that in these constraints, the equality sign holds if cosmic 
strings are responsible for the CMB anistropies.  

The most stringent current limits on $G\mu$, independent of and somewhat 
weaker than structure 
formation considerations, are the gravitational wave constraints based 
on millisecond pulsar timing. The pulsar limits, however, are dominated 
by their sensitivity to the value of $\alpha$. For the values of 
$\alpha$ indicated by the work of Refs. \cite{BB,AS,AT} the pulsar 
limits are consistent with the value of $G\mu / c^2$ needed 
if strings are to provide the seeds for structure 
(see e.g. Refs. \cite{CA,XX} for recent work on this issue). 

As previously stated, in this Letter, we will address the astrophysical 
constraints on black hole formation from 
strings. These bounds can work two ways. For a fixed value of $G\mu$ 
we can constrain the fraction $f$ of string loops which forms black holes. 
Alternatively, given $f$, we can derive constraints on $G\mu$. 
The physical origins of the bounds we analyze are threefold. Firstly, we 
demand that the $\gamma$-ray flux from black holes formed by collapsing 
loops does not exceed the observed $\gamma$-ray background. Secondly, 
we use the observed cosmic ray fluxes (in particular the antiproton flux) 
to limit the contribution of black holes to $\Omega$. Finally, 
exploring the possibility that black holes do not completely vanish 
after their evaporation lifetime but remain as stable Planck scale massive 
relics we can derive additional limits on the string and relic scenarios 
by demanding that black hole relics do not overclose the Universe. 

Cosmic string theory constraints based on the possibility that 
string loops will collapse to black holes were first studied by 
Hawking$^{\cite{SH}}$ and by Polnarev and Zembowicz$^{\cite{AP}}$. They 
pointed out that string loops which collapse under their own tension 
to a radius smaller than their Schwarzschild radius will form black 
holes. For example, a planar circular string loop after a quarter period 
will collapse to a point and hence form a black hole. 
On the other hand, a typical string loops emerging either immediately 
after a phase transition in the early Universe or subsequently 
at a much later time due to the intersection of long strings will 
mostly be asymmetric, and hence have a tiny, but still significant, 
probability of collapsing to a 
radius small enough to form a black hole. We will denote by $f$ the 
fraction of loops produced during the scaling epoch which collapse to 
form 
black holes. Initial estimates$^{\cite{SH,AP,HM}}$ of the constant 
$f$ differed widely. The most recent estimate of $f$ 
is due to Caldwell and Casper$^{\cite{CC}}$ who found, by numerically 
simulating loop fragmentation and evolution, that the fraction 
of loops which collapse to form a black hole within the first 
oscillation period of the loop is
\be \label{fvalue}
f = 10^{4.9 \pm 0.2} (G \mu / c^2)^{4.1 \pm 0.1} \, ,
\ee 
for $G\mu / c^2$ in the range $10^{-3} \alt G \mu / c^2 \alt 3 
\times 10^{-2}$. Caldwell and Casper argue that this parameterization 
can be extrapolated down to $G \mu / c^2 \simeq 10^{-6}$.  

While the amplitude of the mass spectrum of black holes formed by string 
loops collapse is unknown in the cosmic string model, the spectral shape 
is determined by the scaling argument for strings. The number 
density $dn_l / dt$ of string loops formed per unit time can be 
determined from the conservation of string energy
\be \label{masscons}
{\dot \rho_{\infty}} - 2 H \rho_{\infty} = - {{dn_l} \over {dt}} \alpha 
\mu t \, ,
\ee
where $\rho_{\infty} \sim \nu \mu t^{-2}$ is the energy density in the 
long string network. Since a fixed fraction $f$ of these loops forms 
black holes, and the collapse has the greatest probability of occuring 
on the first oscillation of the loop$^{\cite{SH}}$, we obtain
\be \label{inbhdens}
{{dn^p_{BH}(t)} \over {dt}} \propto t^{-4} 
\ee
for the number density (in physical coordinates) of black holes forming 
per time interval. The initial mass $M$ of the black hole formed at time 
$t$ is given by (\ref{inmass}). 

If we neglect the mass loss of black holes with initial mass greater 
than the 
formation  mass of a black hole just expiring today $M_*$ (we later 
justify this to be a good approximation), the present 
number density $dn_{BH} / dM$ of black holes with mass $M$ can be 
obtained by  redshifting the distribution (\ref{inbhdens}) from the 
time of formation $t(M) \agt t(M_*)$ to the present time $t_0$. 
For black holes formed during the radiation-dominated epoch, we 
thus have
\be \label{bhspectrum}
{{dn_{BH}(M)} \over {dM}} \propto M^{-2.5} \,\,\,\, , M_* \alt M.
\ee

The constraints on black hole formation from cosmic strings set 
by the observed $\gamma$-ray background at $100MeV$ have been 
reanalyzed most recently by Caldwell and Gates$^{\cite{CG}}$ and by
Caldwell and Casper$^{\cite{CC}}$. In this Letter, we review and 
correct the analytic arguments in Ref. \cite{CG}, and so derive 
a stronger limit on $G\mu / c^2$. We also make use of the newly 
published EGRET data$^{\cite{EGRET}}$ on the observed 
$30 MeV - 120 GeV$ extragalactic $\gamma$-ray flux. Further 
constraints on black hole formation 
from cosmic strings are then considered. We first reanalyze the 
limits from the observed $\gamma$-ray background. Second, 
we determine the limits from the cosmic ray fluxes. 
Finally, we explore the bounds which can be derived under the 
postulation that black hole relics form. 

\section{Gammay-Ray Flux Constraints} 
 
It is well known$^{\cite{car75,PH,car76}}$ that the extragalactic 
$\gamma$-ray flux observed at $100MeV$ provides a strong constraint 
on the population of 
black holes evaporating today. Too many black holes would lead to an 
excess of such radiation above the observed value. 

In particular, it was shown that if the present day number density 
distribution of black holes of mass $M$ has the form$^{\cite{car75}}$
\be \label{massdistrib}
\frac{dn}{dM} = (\beta -2) \lb\frac{M}{M_{*}}\rb^{-\beta} M_{*}^{-2} 
\Omega_{PBH} \rho_{crit}, \hspace{1.5cm} M_* \alt M  \, ,
\ee     
where $\beta = 2.5$ for black holes formed in the radiation-dominated 
era, $M_*$ is the mass of a black hole whose 
lifetime is the present age of the Universe, $t_o$, and $\Omega_{PBH}$ 
is the present fraction of the critical density in primordial 
black holes, then comparing the Hawking emission from the black 
hole distribution with the $\gamma$-ray background observed by 
the SAS-2 satellite$^{\cite{Fichtel}}$ requires that$^{\cite{mca}}$ 
\be \label{omegalimit}                                                  
\Omega_{PBH} \alt 8 \times 10^{-9} h^{-2}
\ee 
where $h \approx 0.5 - 1.0$ is the Hubble parameter in units of 
$100\; \mbox{km s}^{-1} \mbox{Mpc}^{-1}$. (Note that this limit was 
incorrectly interpreted in Eq. (4.1) of \cite{CG}.) The newly 
published $30 MeV - 120 GeV$ extragalactic $\gamma$-ray 
background measured$^{\cite{EGRET}}$ by the EGRET experiment 
implies an updated limit on the present black hole density 
of$^{\cite{MC97}}$ 
\be \label{newomegalimit}
\Omega_{PBH} \alt (5.1 \pm 1.3) \times 10^{-9} h^{-1.95 \pm 0.15} \, .
\ee
The $\gamma$-ray flux per unit energy from the black hole 
distribution (\ref{massdistrib}) turns$^{\cite{mca}}$ over from an 
$E^{-1.3}$ slope 
below about $10 MeV$ to a steeper slope around $E \simeq 100 MeV$, 
the peak energy of the instantaneous emission from a black hole with mass 
$M_*$. The emission from the black hole distribution falls off as 
$E^{-3}$ above about $1 GeV$. The origin of the observed 
keV - GeV extragalactic $\gamma$-ray background is unknown. 
Since the observed $\gamma$-ray background falls off$^{\cite{EGRET}}$ as 
$E^{-2.10 \pm 0.03}$ between $30 MeV$ and $120 GeV$, 
this raises the possibility that black hole emission may$^{\cite{MC97}}$ 
explain, or contribute significantly, to the observed extragalactic 
background between about $50 - 200 MeV$. 

The distribution of black holes given by (\ref{massdistrib}) is 
precisely the distribution (\ref{bhspectrum}) predicted by the 
cosmic string model. (Such a distribution will also be produced if the 
black holes form in the early Universe from 
scale-invariant adiabatic density perturbations$^{\cite{car75,mca}}$.) 

The fraction of the critical density today in black holes created 
from collapsing cosmic string loops is$^{\cite{CG}}$ 
\be \label{omegaPBH}
\Omega_{PBH}(t_o) = \frac{1}{\rho_{crit}(t_o)} 
\int_{max(t_i,t_*)}^{t_o} dt' \frac{dn_{BH}}{dt'} m(t',t_o) \, ,
\ee 
where the integral is over the time $t'$ when the black holes formed, 
$\frac{dn_{BH}}{dt'}$ is the comoving number density of black holes 
created from loops at time $t'$ and $m(t',t_o)$ is 
the mass at the present time $t_o$ of a black hole formed at time $t'$. 
In (\ref{omegaPBH}), $t_i$ is the formation time of a black hole 
whose initial Schwarzschild radius is equal to the string thickness 
(the minimum initial radius possible from loop collapse),
\be \label{initime}
t_i = \alpha^{-1} \lb\frac{G \mu}{c^2}\rb^{-\frac12} t_{pl}, 
\hspace{1.0cm} t_{pl} = \lb\frac{G \hbar}{c^5}\rb^\frac12  
\ee     
and $t_*$ is the formation time of a black hole with initial mass 
$M_* = \alpha \mu c t_*$, i.e. 
\begin{eqnarray} \label{tstar}
t_* &=& \alpha^{-1} \mu^{-1} c^{-1} M_* \nonumber \\ 
&=& G c^{-3} \alpha^{-1} \lb\frac{G \mu}{c^2}\rb^{-1} M_* \\ 
&\gg& t_i \nonumber
\end{eqnarray}
For an $\Omega = 1$ Universe, 
\be \label{mstar}
M_* \approx 4.4 \times 10^{14} h^{-0.3} \hspace{0.3cm} \mbox{gm} \, , 
\ee                                 
independent of the formation scenario (see \cite{mac}). Because black 
holes with initial masses less than or equal to $M_*$ 
will have evaporated by today, the lower limit of the integral 
in (\ref{omegaPBH}) is $max(t_i,t_*) = t_*$. 
 
We can approximate $m(t',t_0)$ 
by the initial mass of the black hole as given in (\ref{inmass}): 
\be \label{massapprox}
m(t',t_o) \approx \alpha \mu c t' \, .
\ee 
Since black holes with initial masses greater than $M_*$ will 
have evaporated little by today, the approximation (\ref{massapprox}) 
adds an uncertainty of
less than $6\%$ to the value of $\Omega_{PBH}$ in (\ref{omegaPBH}). 
This can be shown by taking the mass loss rate of an individual 
black hole$^{\cite{mac,pag}}$
\be \label{massloss}
\frac{dM}{dt} \approx 5.34 \times 10^{25} \phi(M) M^{-2} 
\hspace{0.3cm} \mbox{gm} \; \mbox{sec}^{-1} \, ,
\ee
solving for $m(t',t_0)$, and comparing a numerical evaluation of the 
resulting integral (\ref{omegaPBH}) with that obtained from the 
analytical approximation using (\ref{massapprox}). In 
(\ref{massloss}), $\phi(M)$ is a slowly increasing function 
which depends on the number of particle species emitted by the 
black hole. $\phi(M)$ is normalized to unity when only massless particles 
are emitted, and $\phi(M_*) \simeq 2$. 
Note that although the mass approximation to be used in (\ref{omegaPBH}) 
is also implicit in (\ref{massdistrib}), the limit on 
$\Omega_{PBH}(t_o)$ in (\ref{omegalimit}) was derived using the 
exact evolved present day spectrum of black holes masses. 

If $f$ is the fraction of cosmic string loops which collapse to form 
black holes in the first period of oscillation, then the number 
density of black holes formed from loops at time $t$ 
is $dn_{BH}/dt = f dn_l/dt$. The assumption that the black hole forms 
on the first loop oscillation or not all has been shown in 
\cite{SH} to be a good approximation. From (\ref{masscons}) and 
({\ref{inbhdens}) it then follows that the number of black holes 
created in a volume $V(t)$ at time $t$ is 
\be \label{bhdistrib}
\frac{dn_{BH}}{dt} = 4 f \frac{A}{\alpha} c^{-3} t^{-4} {{a(t)^3} \over 
{a(t_0)^3}} \, ,  
\ee     
where $A$ is proportional to the number $\nu$ of long strings per Hubble 
volume in the scaling solution, adopting the notation of Ref. 
\cite{CG}, and is found from numerical simulations$^{\cite{BB,AS,AT}}$. 
Because the cosmological scale-factor $a(t)$ is proportional to 
$t^\frac12$ in the radiation-dominated era, and proportional to 
$t^\frac23$ in the matter-dominated era, the integral in (\ref{omegaPBH}) 
is dominated by the black holes created about the time $t_*$ in the 
radiation-dominated era. Noting also that $t_* \ll t_{eq}$ for 
$G\mu / c^2 \agt 10^{-18}$, we have from (\ref{omegaPBH}) 
\begin{eqnarray} \label{omegares1}
\Omega_{PBH}(t_o) &=& \frac{1}{\rho_{crit}(t_o)} \frac{1}{a^3(t_o)} 
\int_{t_*}^{t_o} dt' 4 f A c^{-2} t'^{-3} \mu a^3(t') \nonumber \\  
&\approx& \frac{4 f A \mu c^{-2}}{\rho_{crit}(t_o)} 
\lb\frac{t_{eq}}{t_o}\rb^2 \lb\frac1{t_{eq}}\rb^\frac32 
\int_{t_*}^{t_{eq}} dt' t'^{-\frac32} \\ 
&\approx& \frac{8 f A}{G \rho_{crit}(t_o)} \lb\frac{G \mu}{c^2}\rb 
\lb\frac{t_{eq}}{t_o}\rb^2 \lb\frac1{t_{eq}}\rb^\frac32 
t_{*}^{-\frac12} \nonumber 
\end{eqnarray} 
where$^{\cite{Olive}}$ 
\be
t_{eq}  \approx 3.2 \times 10^{10} \, h^{-4} sec \, .
\ee 
     
Substituting for the critical density of the Universe today 
$\rho_{crit}(t_o)$ 
and the age of the Universe $t_o$ 
\[
\rho_{crit}(t_o) = \frac{3 H_o^2}{8 \pi G} 
\] 
\[ 
t_o = \frac{2 H_o^{-1}}{3} \, ,
\]                                  
for an $\Omega = 1$ Universe, (\ref{omegares1}) becomes
\begin{eqnarray} \label{omegares2}
\Omega_{PBH}(t_o) &\approx& 48 \pi f A \lb\frac{G \mu}{c^2}\rb 
\lb\frac{t_{eq}}{t_*}\rb^\frac12 \\  
&\approx& 5.7 \times 10^{11} \lb\frac{A}{10}\rb f \alpha^\frac12 
\lb\frac{M_*}{4.4 \times 10^{14} \, h^{-0.3} \hspace{0.2cm} 
\mbox{gm}}\rb^{-\frac12} 
\lb\frac{G \mu / c^2}{1.7 \times 10^{-6}}\rb^\frac32 h^{-1.85} \, .
\nonumber 
\end{eqnarray}
Since this must be less than or equal to the observational constraint 
on $\Omega_{PBH}$ given by (\ref{newomegalimit}) for an $\Omega = 1$ 
Universe, then 
\begin{eqnarray} 
f & \alt & 8.9 (\pm 2.3) \times 10^{-21} \lb\frac{A}{10}\rb^{-1} 
\alpha^{-\frac12} 
\lb\frac{M_*}{4.4 \times 10^{14} \, h^{-0.3} \hspace{0.2cm} 
\mbox{gm}}\rb^\frac12 
\nonumber \\ & & \times \lb\frac{G \mu / c^2}{1.7 
\times 10^{-6}}\rb^{-\frac32} h^{-0.1 \pm 0.15} 
\lb {{t_{eq}} \over {3.2 \times 10^{10} \, h^{-4} \hspace{0.2cm} 
\mbox{sec} }} \rb^{-1/2} \, . \label{fconstr1}
\end{eqnarray}
     
Because string loops are predominantly formed from the small-scale 
structure on long strings, the initial loop radius and hence the 
value of $\alpha$ are determined by the physics which sets the scale 
for the small-scale structure. Ref. \cite{ACK} provides evidence 
that this scale may be determined by the strength of the gravitational 
radiation from string loops. If this is so, then we would have the 
relation $\alpha = \gamma G \mu / c^2$, where $\gamma$ is a 
dimensionless coefficient describing the strength of gravitational 
radiation generated by string loops$^{\cite{VaVil}}$, and 
(\ref{fconstr1}) becomes 
\begin{eqnarray} 
f & \alt & 6.8 (\pm 1.7) \times 10^{-19} \lb\frac{A}{10}\rb^{-1} 
\lb\frac{\gamma}{100}\rb^{-\frac12} 
\lb\frac{M_*}{4.4 \times 10^{14} \, h^{-0.3} \hspace{0.2cm} 
\mbox{gm}}\rb^\frac12 
\nonumber \\ & & \times \lb\frac{G \mu / c^2}{1.7 
\times 10^{-6}}\rb^{-2} h^{-0.1 \pm 0.15} 
\lb {{t_{eq}} \over {3.2 \times 10^{10} \, h^{-4} \hspace{0.2cm} 
\mbox{sec} }} \rb^{-1/2} \, . \label{fconstr2} 
\end{eqnarray}          
This is significantly stronger than the Caldwell and 
Gates$^{\cite{CG}}$ limit (4.1) of 
$f \alt 10^{-17}$. The main reason for the difference is that in Ref. 
\cite{CG} the limit of Ref. \cite{mac} was applied incorrectly. 
In order to also compare our new limit to that of Ref. \cite{CC}, we 
rewrite (\ref{fconstr1}) in the same form as Eq. (5.1) 
of \cite{CC}: 
\begin{eqnarray} 
f & \alt & 2.0 (\pm 0.5) \times 10^{-30} \lb\frac{A}{10}\rb^{-1} 
\lb \frac{G\mu}{c^2}\rb^{-2} 
\lb\frac{\gamma G \mu / c^2}{\alpha}\rb^\frac12 
\lb\frac{\gamma}{100}\rb^{-\frac12} \nonumber \\ & & \times
\lb\frac{M_*}{4.4 \times 10^{14} \, h^{-0.3} 
\hspace{0.2cm} \mbox{gm}}\rb^\frac12 
h^{-0.1 \pm 0.15} 
\lb {{t_{eq}} \over {3.2 \times 10^{10} \, h^{-4} \hspace{0.2cm} 
\mbox{sec} }} \rb^{-1/2} \label{fconstr3} \, . 
\end{eqnarray}                  
Thus, our bounds on $f$ are an order of magnitude stronger than those 
of Caldwell and Casper.  
 
So far, we have viewed the constraint on $f$ as an upper bound on the 
admissible value of $f$ given any value of $G\mu$. Conversely, if we 
assume the validity of expression (\ref{fvalue}) for $f$ which was 
derived from the numerical simulation of cosmic string evolution, we can 
deduce an upper bound on the value of $G\mu$. Combining 
(\ref{fconstr3}) and (\ref{fvalue}) yields  
\begin{eqnarray} 
\bigl({{G \mu} \over {c^2}} \bigr)^{6.1 \pm 0.1} \, &<& \, 
10^{-34.6 \pm 0.3} \bigl({{A} \over {10}}\bigr)^{-1} \bigl 
({{\gamma G \mu} \over {\alpha c^2}}\bigr)^{1/2}  \bigl({{\gamma} 
\over {100}}\bigr)^{-1/2} \bigl({{M_*} \over {4.4 \times 10^{14} \, 
h^{-0.3} \hspace{0.2cm} \mbox{gm}}}\bigr)^{1/2} \nonumber \\ & & 
\times \bigl({{\Omega_{PBH}} \over {(5.1 \pm 1.3) \times 10^{-9} \, 
h^{-1.95 \pm 0.15}}}\bigr) \lb {{t_{eq}} \over {3.2 \times 10^{10} \, 
h^{-4} \hspace{0.2cm} \mbox{sec} }} \rb^{-1/2} \, .
\end{eqnarray}       
For $\Omega_{PBH}$ as given by the observed $\gamma$-ray 
background, this requires
\be \label{gmuconstr}
G \mu / c^2 \, \leq \, 2.1(\pm 0.7) \times 10^{-6} \, ,
\ee 
which is in remarkable agreement with the 
bounds$^{\cite{BBS,LP93,ACDKSS}}$ 
on $G\mu / c^2$  obtained by normalizing 
the cosmic string model according to the CMB anisotropy 
data and quoted in the Introduction. We also note that our results are 
considerably improved compared to the 
original estimates in Refs. \cite{SH,AP,HM}, because we have been able 
to make use of the numerical simulations of Ref. \cite{CC} to better 
determine $f$ as a function of $G\mu / c^2$ and make use of the updated 
observational limits on $\Omega_{PBH}$.   

\section{Constraints from the Cosmic Ray Fluxes}
     
The limit on $\Omega_{PBH}$ stated in (\ref{omegalimit}) was found by 
comparing the $\gamma$-ray emission from a cosmological distribution 
of black holes with the observed diffuse extragalactic $\gamma$-ray 
background 
around 100 MeV. Significant limits on $\Omega_{PBH}$ can also 
be derived by considering the antiproton, electron and positron 
emission from a distribution of black holes \cite{mca,mak}. However, 
these limits are more uncertain because, unlike the $\gamma$-ray 
limit, the $\bar{p}$, $e^-$ and $e^+$ limits depend on 
the degree to which black holes cluster in the Galactic halo 
and on the propagation of charged particles in the Galaxy. 

Since the black hole distribution is dominated by holes created 
before the era of galaxy formation, one would expect the black 
holes to cluster in the Galactic halo along with other Cold Dark 
Matter. This leads to an enhancement in the predicted black 
hole contribution to the local cosmic ray flux. Assuming a 
halo model in which the spatial distribution of black holes is 
proportional to the isothermal density distribution of dark 
matter within the Galactic halo and simulating the diffusive 
propagation of antiprotons in the Galaxy, Maki et al.$^{\cite{mak}}$ 
derive an upper limit on $\Omega_{PBH}$ of
\be \label{crlimit}
\Omega_{PBH} \, < \, 1.8 \times 10^{-9} h^{-4/3}
\ee                
based on antiproton data from the BESS `93 balloon 
flight$^{\cite{Yosh}}$. This value would imply a 
limit on  $f$ in (\ref{fconstr1}), (\ref{fconstr2}) 
and (\ref{fconstr3}) that is stronger by a factor of about 
3 and a corresponding limit on 
$G\mu / c^2$ in (\ref{gmuconstr}) of
\be \label{gmuconstr2}
G\mu / c^2 \, <  \, 1.8(\pm0.5) \times 10^{-6} \, ,   
\ee
(\ref{gmuconstr2}) contains a negligible dependence on $h$. 
We note that the antiproton limit on $\Omega_{PBH}$ (\ref{crlimit}) 
was calculated from the BESS '93 data by employing the same solar 
demodulation parameter which is applicable for demodulating the 
proton flux at 1 AU. It is presently a matter of debate as to whether 
there is a charge asymmetry in the solar modulation of the proton 
and antiproton fluxes. 

Future long-duration 
balloon flights$^{\cite{mak}}$ or observations at solar 
minimum$^{\cite{mit}}$ may allow an order of magnitude greater 
sensitivity to $\Omega_{PBH}$ than stated in (\ref{crlimit}) or, 
alternatively, offer the possibility 
of detection of black hole emission. If no black hole antiprotons 
are detected by the proposed experiments with an order of magnitude 
greater sensitivity to $\Omega_{PBH}$, the  constraint in 
(\ref{gmuconstr2}) would then be
\be     
G\mu / c^2 \, <  \, 1.2(\pm0.4) \times 10^{-6} \, ,   
\ee
which would only be in marginal agreement with the requirements on 
$G\mu$ for the cosmic string model of structure formation. Note, 
however, that the method of constraining $G \mu$ using charged 
particle cosmic ray data is intrinsically less certain that the 
$\gamma$-ray constraints because of the uncertainties in the 
clustering of black holes in the Galaxy, and the propagation and 
demodulation of charged particles in the Galaxy and Solar System.

The limits on $\Omega_{PBH}$ derived$^{\cite{mca}}$ by matching the 
emission from a distribution of black holes clustered in the Galaxy 
to the demodulated interstellar electron and positron fluxes at 
$300 MeV$ are also comparable with, and overlap, the $\gamma$-ray 
limit (\ref{newomegalimit}). The origins of the observed antiproton, 
electron and positron spectra between $1 MeV$ and $1 GeV$ have yet 
to be fully understood. This raises the possibility that black hole 
emission may consistently be contributing significantly to the 
extragalactic $\gamma$-ray background and the 
Galactic antiproton, electron and positron backgrounds around 
$100 MeV$.
                                      
\section{Constraints from Black Hole Remnants} 

We can derive further limits on the parameters for black hole 
production from cosmic string loops by considering the possibility 
that black 
holes evolve into stable massive relics after completing their 
lifetime of emission$^{\cite{mac2}}$. The final state of an expiring 
black hole is unknown. The Hawking 
derivation$^{\cite{SH1,SH2}}$ for black hole emission and all 
semiclassical analyses break down when the black hole mass approaches 
$m_{pl}$. At this 
scale, the size of the black hole is comparable to the Compton 
wavelength associated with its mass, and the timescale of the mass 
loss is comparable to the frequency of the quantum radiation. 
Semiclassical back-reaction studies as well as perturbative quantum 
gravitational effects indicate that in this regime higher derivative 
terms become important in the gravitational action. Such terms 
may$^{\cite{MBT}}$ stabilize black holes 
against collapse at $m_{pl}$ or well before the mass reaches 
$m_{pl}$. (See also \cite{prev,prev2} for earlier work, \cite{cgl} 
for a recent review, and \cite{banks} for a study of nonsingular 
black holes in the context of string theory). In an alternative 
scenario, the black hole terminates in an explosion which leaves 
behind a space-time singularity which then vanishes$^{\cite{SWH77}}$ 
or continues to exist as a massless naked singularity$^{\cite{BdW}}$. 

Let us consider the consequences if the black holes created 
from string loops remain as relics with residual mass 
$M_{relic} \geq m_{pl}$ at the 
end of their evaporation. The present fraction of the critical 
density in black hole relics is then 
\be \label{relic1}
\Omega_{relic} \, = \, {{M_{relic}} \over {m_{pl}}} {{m_{pl}} \over 
{\rho_c(t_0)}}\int_{t_{evap}(t_i)}^{t_0} dt_{evap} {{dn_{evap}} \over 
{dt_{evap}}} \, ,
\ee
where $t_{evap}$ is the time when an individual black hole ceases 
evaporating (i.e. reaches mass $M_{relic}$), and 
$dn_{evap} / dt_{evap}$ is the comoving number density of black 
holes expiring per unit time at time $t_{evap}$. $t_i$ is 
defined in (\ref{initime}). By conservation of number density, the 
comoving number density of black holes expiring at $t_{evap}(t)$ is 
equal to the corresponding comoving number density of black holes 
formed from string loop collapse at time $t$. Hence, 
(\ref{relic1}) becomes 
\be \label{relic2}
\Omega_{relic} \, = \, {{M_{relic}} \over {m_{pl}}} {{m_{pl}} \over 
{\rho_c(t_0)}}\int_{t_i}^{t'(t_0)} dt'{{dn_{BH}} \over {dt'}} \, ,
\ee  
where $t'(t)$ is the time when a black hole expiring at time $t$ 
formed. Since $M_* \gg M_{relic}$, we have to good approximation
$t'(t_0) \simeq t_* \ll t_{eq}$, and the time interval in 
the integration lies entirely in the radiation dominated epoch. 
Making use of (\ref{initime}), (\ref{tstar}) and (\ref{bhdistrib}), 
it follows that
\begin{eqnarray} \label{relic3}
\Omega_{relic} \, & = \,  {{4 \, f \, A \, m_{pl} \,t_{eq}^{1/2}} 
\over {\alpha \, c^4 \, t_0^2 \, \rho_c(t_0)}} {{M_{relic}} \over 
{m_{pl}}} \int_{t_i}^{t'(t_0)} dt' t'^{-5/2} \\
& \simeq \, {{16 \, \pi \, f \, \alpha^{1/2} \, A \, t_{eq}^{1/2}} 
\over {t_{pl}^{1/2}}} {{M_{relic}} \over {m_{pl}}} \bigl( {{G\mu} 
\over {c^2}} \bigr)^{9/4} \, . 
\nonumber
\end{eqnarray}
The limit $\Omega_{relic} \leq 1$ then requires
\be
f \, \leq 2.5 \times 10^{-17} \bigl({A \over {10}}\bigr)^{-1} 
\alpha^{-1/2} 
\bigl( {{m_{pl}} \over {M_{relic}}} \bigr) \bigl( {{G\mu / c^2} 
\over {1.7 \times 10^{-6}}}\bigr)^{-9/4} \; h^2 \lb {{t_{eq}} \over {3.2 
\times 10^{10} \, h^{-4} \hspace{0.2cm} \mbox{sec} }} \rb^{-1/2}\, , 
\ee
or, if $\alpha = \gamma G \mu /c^2$,   
\be \label{bound3}
f \, \leq \, 1.9 \times 10^{-15} \bigl( {A \over {10}} \bigr)^{-1} 
\bigl( {{\gamma} \over {100}} \bigr)^{-1/2} \bigl( {{m_{pl}} \over 
{M_{relic}}} \bigr) \bigl( {{G \mu / c^2} \over {1.7 \times 
10^{-6}}} \bigr)^{-11/4} \; h^2 \lb {{t_{eq}} \over {3.2 \times 
10^{10} \, h^{-4} \hspace{0.2cm} \mbox{sec} }} \rb^{-1/2} \, .
\ee

This bound (\ref{bound3}) is substantially weaker than the 
bound on $f$ derived from the emission constraints 
(see (\ref{fconstr2})). However, in the black hole relic scenario, 
the previous emission 
constraints on $f$ must also still hold. We can thus combine these two 
analyses and find the maximum fraction $\Omega_{relic}$ of the critical 
density which is permitted in black hole relics under the condition 
that the 
cosmic string model does not violate the $\gamma$-ray bounds. Inserting 
the constraint on $f$ from (\ref{fconstr2}) in (\ref{relic3}), we obtain 
\be
\Omega_{relic} \, \leq \, 3.6 (\pm 0.9) 10^{-4} \bigl( {{M_{relic}} 
\over {m_{pl}}} \bigr) \bigl( 
{{M_*} \over {4.4 \times 10^{14} \, h^{-0.3} \hspace{0.2cm} 
\mbox{gm}}}\bigr)^{1/2} \bigl( {{G\mu / c^2} \over {1.7 
\times 10^{-6}}}\bigr)^{3/4} \; h^{-2.1 \pm 0.15} \, .
\ee 
Thus, the bound on the black hole formation efficiency factor $f$ 
given by (\ref{fconstr2}) implies that black hole remnants from 
collapsing loops can only contribute significantly to the dark 
matter of the Universe in the cosmic string scenario of structure 
formation ($G\mu / c^2 \simeq 1.7 \times 10^{-6}$) 
if the black hole remnants have a relic mass larger than about 
$10^3 m_{pl}$. Such relic masses naturally arise for example in the 
$SU(N)/Z_N$ theories of Ref. \cite{CPW} in which quantum hair leads 
to stability. If the black hole formation process is less efficient 
than the equality in (\ref{fconstr2}), the remnant mass required for 
an interesting contribution to the cosmological critical density is 
linearly increased.

\section{Conclusions}

We have reconsidered the astrophysical constraints on black hole 
formation from collapsing cosmic string loops, correcting a 
misinterpretation in \cite{CG}, and compared the results with the 
most recent normalizations of the string model for structure formation 
to the CMB anisotropy data.  We investigated limits arising from the 
latest available $\gamma$-ray and cosmic ray flux observations, 
and constraints stemming from the possible formation of black hole 
remnants. 

The $\gamma$-ray emission constraint on the black hole formation 
efficiency  implies that, for the cosmic string parameters which 
follow from the most recent cosmic string COBE
normalizations$^{\cite{BBS,LP93,ACDKSS}}$ and from numerical 
simulations of cosmic string evolution$^{\cite{BB,AS,AT}}$, the maximal 
fraction of string loops which can collapse to form black holes is about 
$f \simeq 2 \times 10^{-18}$. The $G \mu$ dependence of the formation 
efficiency is 
not known, but taking the best available analysis$^{\cite{CC}}$, 
we obtain a bound on $G\mu$ which is remarkably close to the 
value required in the cosmic string model of structure formation. 
The bounds derived from the cosmic ray antiproton, electron and 
positron fluxes are of similar magnitude, but presently afflicted 
with larger uncertainties.

Our results raise the scenario that if cosmic string are responsible 
for the 
CMB anisotropies on large angular scales, then the Hawking emission 
from black holes created by string loop collapse is contributing  
significantly to the observed diffuse extragalactic $\gamma$-ray 
background and to the observed antiproton, electron and positron 
cosmic ray fluxes around 
$100 MeV$. Conversely, future tighter observational limits on the 
cosmic ray backgrounds, in particular on the antiproton flux below 
$1 GeV$, will imply constraints on $G\mu / c^2$ stronger than 
those presently given by the anisotropies detected by the COBE satellite. 

We have also investigated the constraints which can be derived if 
black holes do not evaporate completely, but instead evolve into 
stable massive remnants at the end of their life. We have found that 
unless the mass of the black hole remnants is larger than 
$10^3 m_{pl}$, these 
remnants will contribute negligibly to the dark matter of the 
Universe, even if the black hole formation rate has the maximal 
value allowed by the $\gamma$-ray flux constraints. A remnant mass 
of $10^3 m_{pl}$, however, can arise naturally in some 
models$^{\cite{CPW}}$ of black hole evaporation. In this case, 
cosmic strings could consistently provide an explanation for the 
origin of cosmological structure, for the dark matter, and for the 
origin of the extragalactic $\gamma$-ray  and Galactic cosmic ray 
backgrounds around $100 MeV$.

\centerline{Acknowledgements}

This work has been supported (at Brown) in part by the US Department 
of Energy under contract DE-FG0291ER40688, Task A, and was performed 
while JHM held a NRC-NASA/JSC Senior Research Associateship. We are 
grateful to G. Badhwar, R. Caldwell, B. Carr, M. Salamon, J. Skibo 
and P. Sreekumar for useful communications. JHM and RHB thank the 
Fermilab Astrophysics Theory Group for hospitality during brief visits.

\end{document}